# The cooperative game theory foundations of network bargaining games[*]

MohammadHossein Bateni[†]    MohammadTaghi Hajiaghayi[‡]    Nicole Immorlica[§]

Hamid Mahini[¶]

## Abstract

We study bargaining games between suppliers and manufacturers in a network context. Agents wish to enter into contracts in order to generate surplus which then must be divided among the participants. Potential contracts and their surplus are represented by weighted edges in our bipartite network. Each agent in the market is additionally limited by a capacity representing the number of contracts which he or she may undertake. When all agents are limited to just one contract each, prior research applied natural generalizations of the Nash bargaining solution to the networked setting, defined the new solution concepts of *stable* and *balanced*, and characterized the resulting bargaining outcomes. We simplify and generalize these results to a setting in which participants in only one side of the market are limited to one contract each. The heart of our results uses a linear-programming formulation to establish a novel connection between well-studied cooperative game theory concepts (such as *core* and *prekernel*) and the solution concepts of *stable* and *balanced* defined for the bargaining games. This immediately implies one can take advantage of the results and algorithms in cooperative game theory to reproduce results such as those of Azar et al. [1] and Kleinberg and Tardos [29] and also generalize them to our setting. The cooperative-game-theoretic connection also inspires us to refine our solution space using standard solution concepts from that literature such as *nucleolus* and *lexicographic kernel*. The nucleolus is particularly attractive as it is unique, always exists, and is supported by experimental data in the network bargaining literature. Guided by algorithms from cooperative game theory, we show how to compute the nucleolus by pruning and iteratively solving a natural linear-programming formulation.

## 1 Introduction

Common wisdom has it that the whole is more than the sum of the parts. In more economic terms, this proverb is a translation of the fact that two cooperative agents are often capable of generating a surplus that neither could achieve alone. For example, the owner of a music studio, together with a music band, can record and sell an album; a publishing house, together with an author, can print and sell a book. The proceeds from the album or the book are the surplus generated by the cooperation of the agents. *Bargaining theory* asks how agents divide this jointly generated surplus.

[*]A preliminary version of this work appears in [3] and a short version will be presented in ICALP 2010.

[†]Department of Computer Science, Princeton University, Princeton, NJ; Email: mbateni@cs.princeton.edu; The author is also with Center for Computational Intractability, Princeton, NJ 08540. He was supported by a Gordon Wu fellowship as well as NSF ITR grants CCF-0205594, CCF-0426582 and NSF CCF 0832797, NSF CAREER award CCF-0237113, MSPA-MCS award 0528414, NSF expeditions award 0832797.

[‡]AT&T Labs–Research, Florham Park, NJ; Email: hajiagha@research.att.com

[§]Northwestern University, Chicago, IL; Email: nickle@eecs.northwestern.edu

[¶]Computer Engineering Department, Sharif University of Technology, Tehran, Iran; Email: mahini@ce.sharif.edu; Most of the work was done while the author was visiting Northwestern University, Chicago, IL. The author was in part supported by Iranian National Elites' Foundation.

In the 1950s, John Nash proposed a solution to this problem for the special case of two agents [36]. His solution, known as the *Nash bargaining solution*, is based on the intuition that, all else being equal, agents will tend to divide the surplus equally. If things are not equal, i.e., if some agents have better outside options than others, then the *net* surplus will be divided equally.[1] Since Nash's result, various economists and computer scientists have extended this solution concept to a networked setting and defined new equilibrium concepts as well [6, 29]. In these settings, each agent has a set of potential contracts, represented by the network. The "outside options" of the Nash bargaining solution are now endogenous to the model and generated by the network structure itself.

We propose looking at the network bargaining game through the lens of *cooperative game theory*.[2] In a cooperative game, sets of agents are mapped to values representing the surplus they alone can generate. A solution then assigns a payoff to each agent. The literature has a rich history exploring various solution concepts, their axiomatic properties, and their computability. By interpreting the network bargaining game in this context, we are able to leverage the valuable tools and concepts of cooperative game theory when studying network bargaining. Not only does this enable us to reproduce previous results (with arguably little effort) and derive these previous results for more general models, but more importantly perhaps, we are introduced to new refined solution concepts such as *nucleolus* and *lexicographic kernel*. These concepts are often arguably more predictive than those previously studied.

**Our Results.** In most prior work on networked bargaining, researchers assume each agent can enter *at most one* contract, i.e., the set of contracts form a *matching*.[3] Additionally, contracts are often assumed to be of equal worth. In this paper, we generalize these models by assigning each agent a capacity constraint and allowing him or her to participate in a number of contracts limited by this capacity. We also allow contracts to generate varying amounts of surplus. We mainly focus our efforts on the important special case of bipartite networks, or networks like the music and literature examples above, in which each agent can be divided into one of two types depending on which type of agents he or she contracts with.

The most basic question in these models is to develop predictions regarding which contracts will form and how much each agent will earn as a result. A successful prediction should be *intuitive*, *computationally tractable*, and *unique*. To this end, Kleinberg and Tardos [29] proposed a solution concept for the matching case called *stable* and a refinement called *balanced* that are natural generalizations of the Nash bargaining solution for the two-player case [36].

We first show how to characterize all stable solutions in our general setting using a linear-programming formulation which is a generalization of one developed by Shapley and Shubik [42] for the matching case. We then introduce a special case of our problem in which one side of the market is severely capacity-constrained. In particular, we assume agents on one side of the market can enter into *at most one* contract each while the other side has general constraints. In this constrained setting, we draw connections between balanced and stable outcomes and the cooperative-game-theoretic notions of *core* and *prekernel*. These notions look for solutions that are stable with respect to deviating coalitions of any size. Unlike the general setting, we prove that the set of stable solutions and the core coincide in this setting, as do the set of balanced solutions and the prekernel. This result is of particular interest as the core and prekernel are axiomatic solution concepts of exponential description size (essentially they require that certain conditions hold for *every* subset of agents), whereas the notions of stable and balanced solutions have inherently polynomial descriptions. These connections allow us to leverage existing results in the cooperative game theory literature (such as those for computation of core, prekernel, nucleolus, and lexicographic kernel) to motivate and derive solutions in our setting.

---
[1] Nash's solution concept is actually defined for a slightly more general problem where the feasible contracting region need not be linear; in the special case described herein, Nash's solution corresponds to other standard notions as well. Nonetheless, we adopt the terminology *Nash bargaining solution* to be consistent with prior work on this topic in the computer science literature.

[2] This was done in the economics literature for the matching game described below [42], but the connections we draw between the computer-science-defined notions and the cooperative-game-theory notions are novel.

[3] An exception is the model of Chakraborty and Kearns [6] where agents make multiple deals, but the utilities are non-linear.

As for leveraging tools from cooperative game theory, the connections we draw imply that the techniques of Faigle, Kern, and Kuipers [20] for finding a point in the prekernel of a game can be adapted to find a balanced solution for our constrained bargaining game. Indeed, the resulting algorithm as well as its analysis is essentially the same as the local dynamics given in Azar et al. [1]. These connections also have implications for the model of Kleinberg and Tardos [29]. In their model, the set of possible contracts is not necessarily bipartite, but instead each agent is restricted to participate in at most one contract. Our aforementioned results regarding stable and balanced solutions can be adapted to this setting. Since the set of stable solutions and core coincide, we are able to characterize all graphs which have at least one stable outcome. Namely, a graph has a stable outcome if and only if the simple maximum-weight matching LP relaxation has integrality gap one. Since the set of balanced solutions and the prekernel coincide, we can obtain the Kleinberg-Tardos result for constructing a balanced outcome in polynomial time using simple and well-known results from the economics literature rather than combinatorial constructs such as posets and Edmonds-Gallai decompositions.

Perhaps more importantly, this connection to cooperative game theory guides us in our search for solution concepts for our game. The set of stable/balanced and core/kernel solutions previously proposed may be quite large and hence not terribly predictive. With the goal of computing a unique and predictive outcome for our game, we propose and motivate the cooperative-game-theoretic notion of *nucleolus* as the "right" outcome. The nucleolus is unique and symmetric in that agents with similar opportunities have similar earnings. It is also supported by economic experiments, as discussed in Section 2.2. Additionally, for the above model, we show that the nucleolus is computationally tractable. We prove this by following an iterative linear-programming-based approach used previously by economists [24, 18, 46, 20] and computer scientists [16] in unrelated contexts. In order to adopt this approach to our setting, we show how to prune the linear programs, creating programs of polynomial size. The main technical difficulty is to prove that this pruning maintains the essential relationships between the iterated linear programs and thus still computes the nucleolus.

**Related work.** The most closely related work to ours is that of Kleinberg and Tardos [29]. That paper defines stable and balanced solutions for the matching case mentioned above. They then give an efficient characterization of stable/balanced solutions based on posets and the Edmonds-Gallai decomposition. Our work re-derives and generalizes some of their results using simple and well-known results from the economics literature. Very recently, Azar et al. [1] show that local dynamics does in fact converge to a stable and balanced solution. Incidentally, the connection that we establish between the solution concepts of prekernel and balanced would immediately imply the same local algorithm via a former result of Stearns [47]; see Section 4.1 for more details. We also learned of two other very recent results: Kanoria et al. [27] addresses the problem of finding a "natural" local dynamics for this game, and Azar, Devanur, Jain, and Rabani [2] also study a special case of our problem through cooperative game theory and propose nucleolus as a plausible outcome of the networked bargaining game. The work of Chakraborty et al. [7] as well as that of Chakraborty and Kearns [6] considers a related problem, in which there is no capacity constraints on the vertices but agents have non-linear utilities. They explore the existence and computability of the Nash bargaining solution as well as the proportional bargaining solution in their setting.

Much recent literature has focused on the computability of various solution concepts in the economics literature. In the noncooperative game theoretic setting, the complexity of Nash and approximate Nash equilibria has a rich recent history [48, 31, 30, 12, 13, 9, 10, 5, 14]. In cooperative-game-theoretic settings, the core of a game defined by a combinatorial optimization problem is fundamentally related to the integrality gap of a natural linear program, as observed in numerous prior work [4, 43, 26, 32, 25]. The computability of the nucleolus has also been studied for some special games [16, 28, 46, 20, 21, 19, 24].

Much of our work leverages existing results in the cooperative game theory literature; these results will be cited as they are used.

## 2 Preliminaries

In the network bargaining game, there is a set $N$ of $n$ agents. For bipartite graphs, the set $N$ is partitioned into two disjoint sets $V_1$ and $V_2$ (i.e., $N = V_1 \cup V_2$ and $V_1 \cap V_2 = \emptyset$) and all edges of the network pair one vertex of $V_1$ with one vertex of $V_2$. Each agent $i \in N$ is assigned a *capacity* $c_i$ limiting the number of contracts in which agent $i$ may participate. For each pair of agents $i, j \in N$, we are given a weight $w_{ij}$ representing the *surplus* of a contract between $i$ and $j$ (a weight of $w_{ij} = 0$ means $i$ and $j$ are unable to contract with each other). The capacities together with the weights jointly define a node-and-edge-weighted graph $G = (N, E)$ where $E = \{(i, j) : w_{ij} > 0\}$, the weight of edge $(i, j)$ is $w_{ij}$, and the weight of node $i$ is $c_i$. Our game is fully defined by this construct.

The *(bipartite) bargaining game* is a (bipartite) graph $G = (N, E)$ together with a set of node capacities $\{c_i\}$ and edge weights $\{w_{ij}\}$. There are two special cases of the above game that we consider separately. The first is the *matching game* in which $c_i = 1$ for all $i \in N$ (note the graph need not be bipartite in the matching game). The matching game was studied by Kleinberg and Tardos [29] in the context of bargaining, as well as many economists in the context of cooperative game theory [28, 18, 46]. The second special case is the *constrained bipartite game* in which the graph $G = (V_1 \cup V_2, E)$ is bipartite and the capacities of all agents on one side of the market are one ($c_i = 1$ for all $i \in V_2$).

### 2.1 Solution Concepts

Our main task is to predict the set of contracts $M \subseteq E$ and the division of surplus $\{z_{ij}\}$ that result from bargaining among agents. We call a set of contracts $M$ *feasible* if each node $i$ is in at most $c_i$ contracts: i.e., for each $i \in N$, $|\{j : (i, j) \in M\}| \leq c_i$. A *solution* $(\{z_{ij}\}, M)$ of a bargaining game is a division of surplus $\{z_{ij}\}$ together with a set of feasible contracts $M$ such that the total surplus generated is divided among the agents involved: i.e., for all $(i, j) \in M$, $z_{ij} + z_{ji} = w_{ij}$, and for all $(i, j) \notin M$, $z_{ij} = 0$. We interpret $z_{ij}$ as the amount of money $i$ earns from contracting with $j$. We also define the aggregate earnings of node $i$ by $x_i = \sum_{j \in N} z_{ij}$ and sometimes refer to the set of earnings $\{x_i\}$ as the *outcome* of our game.

The set of solutions of our game is quite large, and so it is desirable to define a subset of solutions that are likely to arise as a result of the bargaining process. There are two approaches one might take in this endeavor. The first is to generalize the bargaining notions introduced by Nash [36] and later extended to networked settings [29]. The second is to study our game from the perspective of cooperative game theory.

In keeping with the bargaining line of work, we define the *outside option* of an agent $i$ to be the best deal he or she could make with someone outside the contracting set $M$. For a fixed agent $k$ with $(i, k) \in E \setminus M$, the best deal $i$ can make with $k$ is to match $k$'s current worst offer. If $k$ is under-capacitated in $M$, $i$ can offer $k$ essentially 0 and so $i$'s outside option with $k$ would be $w_{ik}$. If $k$ is utilized to capacity, then so long as $i$ offers $k$ at least the minimum of $z_{kj}$ over all $j$ such that $(j,k) \in M$, then $k$ will accept the offer. Generalizing this, we get the following definition.

**Definition 2.1.** The *outside option* $\alpha_i$ of agent $i$ in solution $(\{z_{ij}\}, M)$ is

$$\max_{k:(i,k)\in E\setminus M} \max_{j:(j,k)\in M} (w_{ik} - I_k z_{kj}),$$

where $I_k$ is a zero-one indicator variable for whether $k$ is utilized to capacity. If the set $\{k : (i,k) \in E \setminus M\}$ is empty, we define the outside option of $i$ to be zero. The inner maximization is defined to be $w_{ik}$ if its support set is empty.

Intuitively, if an agent has a deal in which he or she earns less than the outside option, then he or she will switch to the other contract. Hence, we call a solution *stable* if each agent earns at least the outside option.

**Definition 2.2.** A solution is *stable* if for all $(i,k) \in M$, $z_{ik} \geq \alpha_i$, and $\alpha_i = 0$ if $i$ has residual capacity.

Nash [36] additionally argued that agents tend to split surplus equally. If agents are heterogeneous in that they have different outside options, then they will split the net surplus equally. Using the terminology of Kleinberg and Tardos [29], we define a solution to be *balanced* if it satisfies Nash's conditions where outside options are defined according to the network structure.

**Definition 2.3.** A solution is *balanced* if for all $(i,k) \in M$, $z_{ik} - \alpha_i = z_{ki} - \alpha_k$ or equivalently $z_{ik} = \alpha_i + \frac{w_{ik} - (\alpha_i + \alpha_k)}{2}$.

Another approach to refining the set of solutions for our game is to study it from the cooperative game theory perspective. A cooperative game is defined by a set of agents $N$ and a value function $\nu : 2^N \to \Re^+ \cup \{0\}$ mapping subsets of agents to the nonnegative real numbers. Intuitively, the value of a set of agents represents the maximum surplus they alone can achieve. Cooperative game theory suggests that the total earnings of agents in a cooperative game is fundamentally related to the values of the sets in which they are contained. To cast our game in the cooperative game theory terminology, we must first define the value of a subset of agents. We will define this to be the best set of contracts they alone can achieve.

**Definition 2.4.** The *value* $\nu(S)$ of a subset $S \subseteq N$ of agents is the maximum $\sum_{(i,j)\in M} w_{ij}$ over all feasible sets of contracts $M$ such that $i, j \in S$ for all $(i,j) \in M$.

In graph-theoretic terminology, this is simply the maximum weighted $f$-factor[4] of the subgraph restricted to $S$ and can be computed in polynomial time.

Cooperative game theory suggests that each set of agents should earn in total at least as much as they alone can achieve. In mathematical terms, we require that the sum of the earnings of a set of agents should be at least the value of that same set. We additionally require that the total surplus of all agents is fully divided among the agents. These requirements together yield the cooperative game-theoretic notion of the *core* [23, 40, 37].

**Definition 2.5.** An outcome $\{x_i\}$ is in the *core* if for all subsets $S \subseteq N$, $\sum_{i \in S} x_i \geq \nu(S)$, and for the grand coalition $N$, $\sum_{i \in N} x_i = \nu(N)$.

---

[4]Given a graph $G(V, E)$ and a function $f : V \mapsto \mathbb{Z}^{\geq 0}$, an *f-factor* is a subset $F \subseteq E$ of edges such that each vertex $v$ has exactly $f(v)$ edges incident on it in $F$. See West [49] for a discussion and for a polynomial-time algorithm to find an $f$-factor. The approach can be extended to the case where $f(v)$ values are upper bounds on the degrees, and we are interested in finding the maximum-weight solution.

The core may be empty even for very simple classes of games, and it may be hard to test whether it is empty or not [11]. However, for our games, we are able to characterize the set of matching games having a nonempty core and show that all bipartite bargaining games have a nonempty core.

Other solution concepts proposed in the cooperative game theory literature are that of *kernel* and *prekernel* [15]. Unlike the core, the kernel and prekernel always exist. As these concepts are closely related and we only work with the prekernel, we only define the prekernel in this paper.[5] The prekernel is defined by characterizing the power of agent $i$ over agent $j$, and requiring that these powers are in some sense equalized. Intuitively, the *power* of $i$ over $j$ is the maximum amount $i$ can earn without the cooperation of $j$.

**Definition 2.6.** The *power* of agent $i$ with respect to agent $j$ in the outcome $\{x_i\}$ is

$$s_{ij}(x) = \max \left\{ \nu(S) - \sum_{k \in S} x_k : S \subseteq N, S \ni i, S \not\ni j \right\}.$$

The *prekernel* is then the set of outcomes $x$ that satisfy $s_{ij}(x) = s_{ji}(x)$ for every $i$ and $j$.

Although the definition of the prekernel is not completely intuitive, it turns out to be similar to the notion of balanced solutions in certain networked settings. A further refinement of this definition is that of *lexicographic kernel* which is, roughly speaking, a subset of the prekernel that lexicographically maximizes the vector of all $s_{ij}$ values. In some sense, this definition tries to be as impartial as possible to different players. As the *nucleolus* defined below is a more widely accepted solution concept and achieves complete impartiality, we do not give detailed information about the lexicographic kernel. We simply note that it has been studied in [22, 33, 50], and the result of [22] in addition to Lemma 4.6 allows us to compute the lexicographic kernel for any general bipartite (or even nonbipartite) bargaining game.

## 2.2 A unique outcome

None of the solution concepts proposed above are unique. For any given game instance, and any of the solution concepts above, there may be many outcomes which satisfy it. For example, consider a bipartite bargaining game with two vertices on each side. Each of the four possible edges has value one, and the capacities of the vertices are also one. It can be easily verified that any solution assigning value $x$ to the vertices of one side and $1 - x$ to the other side for $0 \le x \le 1$ is a stable, balanced solution, and hence (as we will show later) also in the core and kernel. However, the solution corresponding to $x = 0$ seems in some sense unfair. After all, all agents appear symmetric, as can be formalized by the fact that there exists an automorphism of the game mapping any agent into any other agent. Hence, we expect the agents to have similar earnings after the bargaining procedure. Among all the plausible solution concepts, the one we expect to see is that for which $x = 1/2$, i.e., each agent earns $1/2$. This intuition was tested and verified in the laboratory experiments of Charness et al. [8]. The solution $x = 1/2$ turns out to be the *nucleolus* of the example game. In general, the nucleolus is that outcome which maximizes lexicographically the excess earnings of any given set.

**Definition 2.7.** Given an outcome $\{x_i\}$, the *excess* $\epsilon(S)$ of a set $S$ is the extra earnings of $S$ in $\{x_i\}$: $\epsilon(S) = \sum_{i \in S} x_i - \nu(S)$. Let $\epsilon = (\epsilon_{S_1}, \ldots, \epsilon_{S_{2^N}})$ be the vector of excesses sorted in nondecreasing order. The *nucleolus* of a bargaining game is the outcome which maximizes, lexicographically, this vector $\epsilon$.

---

[5]In fact, kernel and prekernel coincide in our game because $\nu(\{i\}) = 0$ for any $i \in N$—indeed, the two closely-related solution concepts coincide for any *zero-monotonic TU-game* [35], and our game is one of this class.

The nucleolus was first introduced by Schmeidler [41]. It is a point in the kernel [41], and also part of the core if it is nonempty. We will show later that any point in the intersection of core and kernel must be stable and balanced (at least for the matching and constrained games), and hence the nucleolus inherits all the nice properties of stable and balanced solutions. In addition, the nucleolus is unique [17], and is in fact characterized by a set of simple, reasonable axioms [39, 44, 34, 45] including our intuitive notion of symmetry mentioned above.

## 2.3 Results

In this paper, we are primarily interested in developing natural solution concepts for our bargaining game. We posit that such a solution concept should be *intuitive*, *computationally tractable*, and *unique*. Building on prior work, we offer the set of stable solutions as an intuitive solution concept and provide a complete characterization of this set based on a linear-programming interpretation of the bargaining game. The following theorem is proved in Section 3.2.1.

**Theorem 2.8.** *The set of all stable solutions to the network bargaining game can be constructed in polynomial time.*

The set of stable solutions in our game might be quite large, and so, following prior work, we propose balanced solutions as a refinement. We first study the relationship between the stable/balanced concepts and the core/kernel concepts from cooperative game theory. We find that, while these solutions may differ in general, for the constrained bargaining game they exactly coincide. This provides additional motivation for these solution concepts and additionally gives us computational insights from the cooperative game theory literature. We prove the following theorem via Lemmas 3.4 and 3.5. and 3.2.2.

**Theorem 2.9.** *An outcome $\{x_i\}$ of the constrained bipartite bargaining game is in the core if and only if it corresponds to a stable solution $(\{z_{ij}\}, M)$. That is, $\{x_i\}$ is in the core if and only if there exists a stable solution $(\{z_{ij}\}, M)$ such that $x_i = \sum_j z_{ij}$ for all agents $i$.*

Section 3.2.2 proves the following theorem.

**Theorem 2.10.** *An outcome $\{x_i\}$ of the constrained bipartite bargaining game is in the core intersect prekernel if and only if it corresponds to a balanced solution $(\{z_{ij}\}, M)$. That is, $\{x_i\}$ is in the core intersect prekernel if and only if there exists a balanced solution $(\{z_{ij}\}, M)$ such that $x_i = \sum_j z_{ij}$ for all agents $i$.*

The proofs of theorems 2.8, 2.9 and 2.10 can also be adapted to the matching game studied by Kleinberg and Tardos [29], enabling us to recover some of their results using simple and well-known cooperative-game-theoretic constructs; see Section 4.

Using the work of Faigle, Kern, and Kuipers [20], Theorem 2.10 implies an algorithm for computing *some* balanced solution in constrained bipartite bargaining games (as well as the matching game studied by Kleinberg and Tardos [29]). The following theorem is proved in Section 4.1.

**Theorem 2.11.** *There is a local dynamics[6] that converges to an intersection of prekernel and core, which is a stable, balanced solution of the KT game.*

---
[6]In fact, the resulting algorithm is quite similar to the recent local dynamics proposed in Azar et al. [1].

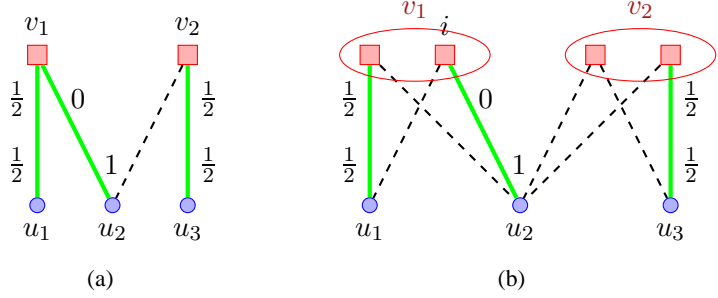

Figure 1: Reducing to matching game model via the copying idea fails.

However, the set of balanced solutions can be quite large, and not all balanced solutions are intuitive. A unique and intuitive balanced solution is the nucleolus, motivated by its symmetry properties. Our final result is an algorithm for computing the nucleolus in constrained bipartite bargaining games. Section 5 is devoted to proving the theorem below.

**Theorem 2.12.** *The nucleolus of the constrained bipartite bargaining game can be computed efficiently.*

## 3 Characterizing Solution Concepts

In this section, we study the solution concepts posed in Section 2 for bipartite graphs with arbitrary capacity constraints. We first define a polytope characterizing all stable solutions (even for non-bipartite graphs). This demonstrates that stable solutions can be computed efficiently and hence helps us understand likely outcomes of our game. We also use our characterization to illustrate the connection between the sets of stable/balanced solutions and the core/kernel of the corresponding cooperative game. This allows us to compute a balanced solution by leveraging existing algorithms for finding a point in the kernel of a cooperative game.

### 3.1 Characterizing Stable Solutions

We begin by characterizing all stable solutions in the general network bargaining game. In order to do this, a natural approach would be to make "ghost" copies of each node, thereby transforming the general case to the matching case. We first demonstrate that this approach does not work even in the constrained bipartite case.

Given a (constrained) bipartite bargaining game $G = (N = V_1 \cup V_2, E)$, we replace each agent $i \in N$ with $c_i$ "ghost" copies. For each $(i, j) \in E$ and each ghost copy of $i$ and ghost copy of $j$, we additionally add an edge of weight $w_{ij}$. Consider the example in Figure 1(a). The agents in the upper row have capacities two, whereas the agents in the bottom row have capacities one. All edges have weight one. The solid edges depict a feasible set of contracts. First, we observe that the given solution is stable. We only consider the agents $u_2$ and $v_2$, since the outside options for the other agents are zero. The profit of the agent $u_2$ is \$1 and her outside option cannot be better than \$1. On the other hand, the outside option of $v_2$ is zero since his only outside option is to make an agreement with $u_2$ whose current option has value \$1. Looking at the transformed instance in Figure 1(b), one can verify that the outside option of agent $i$ is $\frac{1}{2}$ while his profit is zero. Hence, the previously stable solution is no longer stable in this transformed instance.

Therefore, we are unable to use the existing poset-based characterization of Kleinberg and Tardos [29] for the matching case to solve the general case. Instead, we define a linear program describing the set of op-timal contracts and its dual, and use these to characterize the stable solutions, thereby proving Theorem 2.8. Our linear program is a generalization of the one used by Shapley and Shubik [42] to describe the core for the simpler matching version of the network bargaining game. The optimal contracts can be described by the following linear program:

$$
\begin{aligned}
\text{maximize} \quad & \sum_{ij} w_{ij} x_{ij} \\
\text{subject to} \quad & \sum_j x_{ij} \leq c_i & \forall i \in N \\
& 0 \leq x_{ij} \leq 1 & \forall (i,j) \in E(G),
\end{aligned}
\quad \text{(LP1)}
$$

where there is a variable $x_{ij}$ for every edge $(i,j) \in E$. The dual of LP1 is:

$$
\begin{aligned}
\text{minimize} \quad & \sum_i u_i c_i + \sum_{ij} y_{ij} \\
\text{subject to} \quad & u_i + u_j + y_{ij} \geq w_{ij} & \forall (i,j) \in E(G) \\
& 0 \leq y_{ij} \leq 1 & \forall (i,j) \in E(G) \\
& 0 \leq u_i \leq 1 & \forall i \in N.
\end{aligned}
\quad \text{(LP2)}
$$

Given an optimal pair of solutions to these LPs, we show how to construct a stable solution. The primal variables indicate the set of optimal contracts $M$. We use the dual solution to divide the surplus $w_{ij}$ of the contracts in $M$. For each contract $(i,j) \in M$, we give $u_i$ to $i$, $u_j$ to $j$, and then divide arbitrarily the remaining $y_{ij}$. Thus $z_{ij} = u_i + \alpha_{ij} y_{ij}$ and $z_{ji} = u_j + (1 - \alpha_{ij}) y_{ij}$ for an arbitrary $\alpha_{ij} \in [0,1]$ (different $\alpha_{ij}$ yield different stable solutions). Conversely, to convert any stable solution to a pair of optimal solutions, we set the primal variables based on the contract $M$. We define the dual variables as follows: (1) for every unsaturated vertex $i$, set $u_i = 0$; (2) for every saturated vertex $i$, set $u_i = \min_{j \in N_i} z_{ij}$; (3) for every $x_{ij} = 0$, set $y_{ij} = 0$; (4) for every $x_{ij} = 1$, set $y_{ij} = w_{ij} - u_i - u_j$. To prove that these constructions work, we use the complementary slackness conditions (see Lemmas 3.1 and 3.2 below):

1. $(c_i - \sum_j x_{ij})u_i = 0$,

2. $(1 - x_{ij})y_{ij} = 0$, and

3. $(u_i + u_j + y_{ij} - w_{ij})x_{ij} = 0$.

## 3.2 Relating bargaining solutions to cooperative game theory solutions

Whereas the notions of stable and balanced solutions have been only recently introduced, the cooperative game theoretic notions of core and kernel are well-studied. Hence it is interesting to relate these two seemingly different notions. There is no general reason why we would expect these notions to coincide, and in fact, as we demonstrate via an example below, they do not for the general bipartite bargaining case. Nonetheless, for the constrained bipartite bargaining game, we are able to prove that stable/balanced and core/kernel do coincide. In the discussions below, we say outcome $\{x_i\}$ has been produced by solution $(\{z_{ij}\}, M)$ if $x_i$ is equal to total amount of money that agent $i$ earns in solution $(\{z_{ij}\}, M)$.

### 3.2.1 The general bipartite bargaining games

Consider the bargaining game in Figure 2, with the depicted outcome vector $\{x_i\}$ consisting of $0.8$ and $1.2$ values. All the contract values are one, and the capacities of the agents are two. Let us first see that this cannot correspond to a stable solution. The grand coalition has value 8, and the edge $(v_2, u_3)$ cannot participate in any maximum feasible set of contracts. Consider the agents $v_2$ and $u_3$. Each has a contract from which she earns less than $0.5$. So, they have the incentive to cancel those contracts and sign the contract $(v_2, u_3)$ between themselves. Hence, this is not a stable outcome.

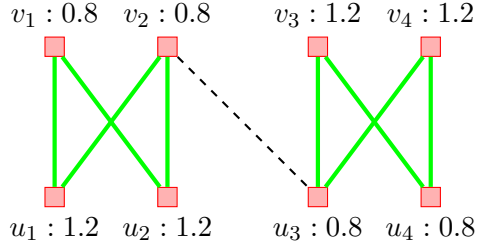

Figure 2: Core $\not\subseteq$ Stable in unconstrained bipartite bargaining games.

We now prove that the vector $\{x_i\}$ is indeed in the core of the game. The total surplus (8 dollars) is all distributed among the agents. Coalitions of size zero or one are satisfied because their game value is zero. No coalition of size two has value more than one, whereas they have secured at least $2 * \$0.8 = \$1.6$. Coalitions of size three secure at least \$2.4 while their game value is at most two. A coalition of size four cannot have value larger than four (one side has at most two vertices and the capacities are two. However, any four agents obtain at least \$3.2, and if the game value is indeed larger than three, it has to be either of the left or right four green edges, in which case the four vertices are guaranteed four dollars. Coalitions of size 5 and 7 have values at most 4 and 6, respectively (by the same argument as the above). Such coalitions receive \$4.4 and \$6.8, respectively. This leaves us to consider coalitions of size six. They are guaranteed to receive \$5.6 while their game value is no more than 5; since to have more than five feasible contracts, both sides should contain three vertices, and it is clear two vertices cannot exhaust their capacities.

The above example shows that core is not necessarily contained in the set of stable solutions, however, we show below that for the general bipartite bargaining games, the converse is true; i.e., the set of stable solutions is indeed a subset of the core of the game.

We first show how to use solutions to LP1 and LP2 to produce stable solutions. Later on, we prove that all stable solutions can be obtained in this way. The following two lemmas imply Theorem 2.8.

**Lemma 3.1.** *For every integer optimum solution $X$ of LP1 and every real optimum solution $Y$ of LP2, we can construct a set of stable solutions $\{(\{z_{ij}\}, M)\}$.*

**Proof.** Consider an integer optimum solution $X$ of LP1 and a real optimum solution $Y$ of LP2. For every $x_{ij} = 1$, there will be a contract between $i$ and $j$. Hence $M = \{(i,j) : x_{ij} = 1\}$.

To set $z_{ij}$, we use the dual solution $Y$. We first set $z_{ij} = u_i$ for every $i$ and $j$ with $x_{ij} = 1$ and $z_{ij} = 0$ for every $i$ and $j$ with $x_{ij} = 0$. We then divide $y_{ij}$ arbitrarily between $z_{ij}$ and $z_{ji}$. Thus $z_{ij} = u_i + \alpha_{ij} y_{ij}$ and $z_{ji} = u_j + (1-\alpha_{ij}) y_{ij}$ for an arbitrary $\alpha_{ij} \in [0,1]$ (different $\alpha_{ij}$ yield different stable solutions).

Note that, by complementary slackness, $(u_i + u_j + y_{ij} - w_{ij}) x_{ij} = 0$. So $u_i + u_j + y_{ij} = w_{ij}$ for every $x_{ij} = 1$. This shows we divide exactly $w_{ij}$ between $i$ and $j$ for every $x_{ij} = 1$. In order to prove stability we should prove that no vertex has a better outside option. By construction any vertex $i$ gets at least $u_i$ in each contract. We show $i$ cannot get more than $u_i$ from an outside option. Assume $j$ is an outside option for $i$. Thus $x_{ij} = 0$, and so the complementary slackness condition $(1 - x_{ij}) y_{ij} = 0$ implies that $y_{ij} = 0$. On the other hand, by feasibility of the dual, $u_i + u_j + y_{ij} \geq w_{ij}$. Thus $u_i \geq w_{ij} - u_j$. If $j$ is not saturated (i.e., $j$ has enough capacity to enter into another contract), then complementary slackness condition $(c_i - \sum_j x_{ij}) u_i = 0$ implies that $u_j$ is zero and so $u_i \geq w_{ij}$ meaning that $i$'s current options are superior to a contract with $j$. Alternatively, if $j$ is saturated, then $j$ receives at least $u_j$ in each of his contracts by construction and so $i$ can earn at most $w_{ij} - u_j$ in a contract with $j$. Again, as $u_i \geq w_{ij} - u_j$, $i$'s current options are superior to a contract with $j$, and so our solution is stable. □

For every integer solution $X$ and real solution $Y$ of LP1 and LP2 we have constructed a set $S_{X,Y}$ of stable solutions (one for each possible choice of $\alpha_{ij}$'s). In the following lemma we prove the set $S = \bigcup_{X,Y} S_{X,Y}$ is equal to the set of all stable solutions.

**Lemma 3.2.** *For every stable solution $(\{z_{ij}\}, M)$, there exists an integer optimum solution $X$ of LP1 and real optimum solution $Y$ of LP2 for which $(\{z_{ij}\}, M) \in S_{X,Y}$.*

**Proof.** Assume $(\{z_{ij}\}, M)$ is a stable solution of our game. If there is no agreement between $i$ and $j$ then $z_{ij} = z_{ji} = 0$. If there is an agreement between $i$ and $j$ then $z_{ij} + z_{ji} = w_{ij}$. First we construct an integer solution of LP1 by setting $x_{ij} = 1$ if and only if $(i, j) \in M$. Next we construct a real solution of LP2. First define $N_i$ to be the set of vertices $j$ such that $(i, j) \in M$. We say $i$ is saturated if $|N_i| = c_i$. Now we define the real solutions of LP2 as follows: (1) for every unsaturated vertex $i$, set $u_i = 0$; (2) for every saturated vertex $i$, set $u_i = \min_{j \in N_i} z_{ij}$; (3) for every $x_{ij} = 0$, set $y_{ij} = 0$; (4) for every $x_{ij} = 1$, set $y_{ij} = w_{ij} - u_i - u_j$ (note that $z_{ij} + z_{ji} = w_{ij}$ in this case, and $u_i \leq z_{ij}$, and $u_j \leq z_{ji}$, so $u_i + u_j \leq w_{ij}$ and $y_{ij}$ will be non-negative).

We prove the above solutions are optimum solutions for LP1 and LP2 using complementary slackness. In the first and second steps we set $u_i$'s such that $(c_i - \sum_j x_{ij})u_i = 0$ holds. In the third step we set $y_{ij}$ such that $(1 - x_{ij})y_{ij} = 0$ holds. By definition of $y_{ij}$ in the last step, we have $(u_i + u_j + y_{ij} - w_{ij})x_{ij} = 0$. Using complementary slackness, we can conclude that both of our solutions are optimal for LP1 and LP2. □

Implicit in the above construction is that any stable solution corresponds to a maximum-weight $f$-factor in the graph. With slight abuse of notation, we redefine an $f$-factor as any subset $E'$ of edges of the graph in which $|\{u|(u, v) \in E'\}| \leq f(v)$ for all $v \in V(G)$. Notice that in the conventional definition, the degree condition holds with equality.

**Corollary 3.3.** *A network bargaining game has a stable solution if and only if LP1 has integrality gap one for finding a maximum-weight $f$-factor.*

It is well-known that the condition holds true for all bipartite graphs [38]. The condition however can be verified for any given graph to test whether the given instance has any stable solutions or not.

**Lemma 3.4.** *For every integer optimum solution $X$ of LP1 and every real optimum solution $Y$ of LP2, all the vectors in $S_{X,Y}$ are in the core.*

**Proof.** Consider an arbitrary element $(\{z_{ij}\}, M)$ of $S_{X,Y}$ and assume $y_{ij} = y'_{ij} + y'_{ji}$ has been divided between $i$ and $j$ such that $i$ gets $y'_{ij}$ and $j$ gets $y'_{ji}$. Player $i$ earns $x_i = c_i u_i + \sum_j y'_{ij}$ from all her contracts since (1) she earns $u_i + y'_{ij}$ from her contract to $j$, and (2) she has exactly $c_i$ contracts if $u_i \neq 0$.

We prove that $\mathbf{x} = (x_1, \ldots, x_n)$ is in the core. First we show that the core condition holds for the grand coalition $N$, i.e., $\sum_i x_i = \nu(N)$:

$$\sum_{i \in N} x_i = \sum_{i \in N} \left( c_i u_i + \sum_{j \in N} y'_{ij} \right) = \sum_{i \in N} c_i u_i + \sum_{i,j \in N} y'_{ij} = \sum_{i \in N} c_i u_i + \sum_{i,j \in N} y_{ij}$$

which is the value of an optimal dual solution, hence by strong duality,

$$= \nu(N).$$

Next consider a set $R \subset N$. If we write LP2 restricted to set $R$, it is clear that values $u_i$ and $y_{ij}$, $i, j \in R$, construct a feasible solution for this linear program. Therefore, their sum is a dual feasible solution, hence by weak duality, at least $\nu(R)$. In other words, we have

$$\sum_{i \in R} x_i = \sum_{i \in R} (c_i u_i + \sum_{j \in N} y'_{ij}) \geq \sum_{i \in R} c_i u_i + \sum_{i,j \in R} y'_{ij} = \sum_{i \in R} c_i u_i + \sum_{i \in V_1 \cap R, j \in V_2 \cap R} y_{ij}$$

which is the value of a feasible dual solution restricted to vertices in $R$, and by weak duality,

$$\geq \nu(R). \qquad \square$$

### 3.2.2 The constrained bipartite bargaining game

Now we focus on the constrained case and show that stable/balanced coincides with core/kernel as promised.

**Stable = Core.** We already proved that an outcome $\{x_i\}$ produced by any stable solution $(\{z_{ij}\}, M)$ is in the core, even for a more general case (i.e., in the general bipartite case). Now we prove the other direction (i.e., Core is a subset of Stable) for the constrained bipartite case. We show that for every $\{x_i\}$, there is a stable solution $(\{z_{ij}\}, M)$ producing $\{x_i\}$. In the constrained bipartite graph $c_j = 1$ for every $j \in V_2$, and thus given the set of contracts $M$, the $\{z_{ij}\}$ are uniquely determined by the core outcome $\{x_i\}$ (for every $(i, j) \in M$ with $j \in V_2$ we set $z_{ji} = x_j$ and $z_{ij} = w_{ij} - x_j$). The crux of the problem is therefore in choosing $M$ and using the core inequalities to prove that the resulting solution is stable.

**Lemma 3.5.** *Every outcome $\{x_i\}$ in the core is produced by some stable solution $(\{z_{ij}\}, M)$.*

**Proof.** Consider any optimal set of feasible contracts $M$. For all $(i, j) \in M$ with $i \in V_1$ and $j \in V_2$, set $z_{ij} = w_{ij} - x_j$ and $z_{ji} = x_j$. We claim the solution $(\{z_{ij}\}, M)$ is stable and produces $\{x_i\}$.

In order to show our solution is feasible, we should prove that for every $j \in V_2$ either $x_j = 0$ or $0 < x_j \leq w_{ij}$ and $(i, j) \in M$. Suppose $x_j > 0$. We know that $\nu(N) = \sum_{k \in N} x_k > \sum_{k \in N \setminus \{j\}} x_k \geq \nu(N \setminus \{j\})$. Thus, $\nu(N \setminus \{j\}) \neq \nu(N)$ implying $j$ participates in all maximum feasible set of contracts. Hence there is some $(i, j) \in M$. If we remove edge $(i, j)$, then we have a feasible solution for $N \setminus \{j\}$. Therefore $\nu(N \setminus \{j\}) \geq \nu(N) - w_{ij}$. On the other hand, we know $\sum_{i \in N} x_i = \nu(N)$ and $\sum_{i \in N \setminus \{j\}} x_i \geq \nu(N \setminus \{j\})$ by core properties, which implies $x_j \leq w_{ij}$. Therefore our solution is feasible.

To see that it is stable, note that the edges in $M$ form a union of stars each of which has exactly one vertex in $V_1$, namely its "center". Denote these stars by $R_1, \ldots, R_m$ and their centers by $r_1, \ldots, r_m$. Let $w(R_k) = \sum_{j \in R_k \cap V_2} w_{r_k, j}$ be the sum of weights of edges in $R_k$. It is clear that $\nu(R_k) = w(R_k)$ and $\nu(N) = \sum_k w(R_k) = \sum_k \nu(R_k)$. We know that $\{x_i\}$ is in the core, so we have: (1) for the grand coalition $N$, $\sum_{i \in N} x_i = \nu(N) = \sum_{k=1}^m w(R_k)$; and (2) for each star $R_k$, $\sum_{i \in R_k} x_i \geq \nu(R_k) = w(R_k)$. From this, we can conclude that for each star, $\sum_{i \in R_k} x_i = w(R_k) = \nu(R_k)$. We prove no vertex in $V_1$ wants to change his contracts. This guarantees stability, since if a vertex of $V_2$ prefers to change her contract to another vertex $v$ in $V_1$, this implies that $v$ should also prefer this new contract to one of her existing contract which is a contradiction. Consider an arbitrary vertex $r_k \in V_1$. We say $r_k$ is saturated if it has exactly $c_{r_k}$ edges in $M$. There are two possibilities. If $r_k$ is not saturated, then consider an edge $(r_k, j) \notin M$. We know that $\sum_{i \in R_k} x_i = w(R_k)$ and $\sum_{i \in R_k \cup \{j\}} x_i \geq \nu(R_k \cup \{j\}) = w(R_k) + w_{r_k, j}$. Therefore $x_j \geq w_{r_k, j}$ and $r_k$ gets zero in any contract with $j$. If $r_k$ is saturated, then consider an edge $(r_k, j) \notin M$ and an edge $(r_k, j') \in M$. We prove $w_{r_k, j'} - x_{j'} \geq w_{r_k, j} - x_j$ and so $r_k$ will not change contracts. We know $\sum_{i \in R_k} x_i = w(R_k)$ and $\sum_{i \in R_k \cup \{j\} \setminus \{j'\}} x_i \geq \nu(R_k \cup \{j\} \setminus \{j'\}) = w(R_k) - w_{r_k j'} + w_{r_k j}$. Therefore $x_j - x_{j'} \geq w_{ij} - w_{ij'}$ implying $w_{ij'} - x_{j'} \geq w_{ij} - x_j$. $\square$

**Balanced = Kernel.** Next, we prove that Stable ∩ Balanced is equivalent to Core ∩ Kernel. From the discussion above, we can assume we are provided with a stable solution $(\{z_{ij}\}, M)$ and associated core outcome $\{x_i\}$. The goal is to show that the solution is balanced if and only if the outcome is in the kernel.

Recall that the outcome $\{x_i\}$ is in kernel if and only if $s_{ij} = s_{ji}$ for all pairs of players $i, j$. On the other hand, a solution is balanced if and only if for any $(i, j) \in M$, $z_{ij} - \alpha_i = z_{ji} - \alpha_j$. We define the net gain[7] of player $i$ after losing the contract with $j$ as $\tilde{s}_{ij} := \alpha_i - z_{ij}$ and prove in the following that $\tilde{s}_{ij} = s_{ij}$ for all edges $(i, j) \in M$. This finishes the proof.

We show that the maximum for $s_{ij}$ occurs necessarily at a set $T$ such that $T = R_{i'} \setminus \{j'\} \cup \{j''\}$ for some $i' \in V_1, j', j'' \in V_2$ or at $T = R_{i'} \setminus \{j'\}$ for some $i' \in V_1, j' \in V_2$, where $R_{i'}$ denotes the star rooted at $i'$. Clearly, $T$ has at least one vertex $i' \in V_1$, since otherwise, $\nu(T) = 0$.

First suppose for the sake of reaching a contradiction that there are at least two vertices of $V_1$ in $T$, say $i' = i_1, i_2, \ldots, i_k$. Let $M'$ be a maximum feasible set of contracts in the subgraph induced by $T$. $M'$ partitions $V_2 \cap T$ into $T_0, T_1, \ldots, T_k$ where $T_l$ for $1 \leq l \leq k$ should sign a contract with $i_l \in V_1$ while $T_0$ will not be saturated. Clearly, we can assume $T_0 = \emptyset$, since removing it from $T$ does not change the value of $\nu(T)$ whereas it may decrease $x(T)$, thus increasing the expression $\nu(T) - x(T)$. Notice that $x(\{i_l\} \cup T_l) \geq \nu(\{i_l\} \cup T_l)$ for all $1 \leq l \leq k$, because $\{x_i\}$ is in the core. Therefore, we get

$$\nu(T) - x(T) = \sum_{1 \leq l \leq k} [\nu(\{i_l\} \cup T_l) - x(\{i_l\} \cup T_l)]$$

since $\nu(T) = \sum_l \sum_{j \in T_l} w_{i_l, j}$ and $\nu(\{i_l\} \cup T_l) = \sum_{j \in T_l} w_{i_l, j}$,

$$\leq \nu(\{i_1\} \cup T_1) - x(\{i_1\} \cup T_1).$$

Hence, we can assume we only have one vertex of $V_1$ in the sets we consider.

Next, assume $i'$ is the only vertex in $V_1 \cap T$. We show that $|T \setminus \{i'\}| \leq c_{i'}$. Since no more than $c_{i'}$ vertices of $T$ can be connected to $i'$ in any feasible set of contracts, there is a set $T' \subseteq T$ such that $|T' \setminus \{i'\}| \leq c_{i'}$ and $\nu(T') = \nu(T)$. Removing the vertices in $T \setminus T'$ does not change $\nu(T)$, and it cannot increase $x(T)$. So, we assume $|T \setminus \{i'\}| \leq c_{i'}$.

Furthermore, we may assume there is only one $j' \in T_{i'} \setminus T$. We add the other vertices one at a time, without decreasing the value $\nu(T) - x(T)$. Take any other vertex $j_1 \in T_{i'} \setminus T$. Add $j_1$ to $T$. If $\nu(T)$ goes up by $w_{i', j_1}$, $\nu(T) - x(T)$ cannot decrease, since $x_{j_1} \leq w_{i', j_1}$. However, if $\nu(T)$ does not increase as much, the number of edges of $i'$ has to stay the same; we then remove the unique vertex $j_2 \in T \setminus T_{i'}$ that was exchanged for $j_1$ in the maximum feasible set of contracts. The stability of the solution guarantees that $w_{i', j_1} - x_{j_1} \geq w_{i', j_2} - x_{j_2}$, thus this step does not decrease $\nu(T) - x(T)$ either.

Having established the structural property, we can now prove that $s_{ij} = \tilde{s}_{ij}$. First assume that $i \in V_1, j \in V_2$. We first note that the definition of $s_{ij} := \max_{T \ni i, T \not\ni j}[\nu(T) - x(T)]$ implies $s_{ij} \geq \tilde{s}_{ij}$, since the latter corresponds to one of the terms over which the maximization is performed: $T = R_i \cup \{j'\} \setminus \{j\}$ if $i$'s outside option is nonzero, and $T = R_i \setminus \{j\}$ otherwise. The restriction on the sets $T$ to consider narrows down the definition to the outside options considered for $\tilde{s}_{ij}$: $i$ has to be included and is the only vertex in $V_1 \cap T$, $j$ has to be removed, and we may add another vertex $j'' \in V_2$.

Next, consider the case of $s_{ji}$. By the above arguments, we restrict ourselves to sets $T \ni i'$ for some $i' \in V_1 \setminus \{i\}$, such that at most one vertex from $R_{i'}$ is removed in $T$, and hence, at most one vertex is added. The latter has to be $j$ itself. Therefore, this precisely models the outside options of $j$ threatened by $i$, and $s_{ji} = \tilde{s}_{ji}$.

---

[7] In fact, it is more intuitive to define the net loss (which is equal to $-\tilde{s}_{ij}$ in our notation), however, we define the gain to be consistent with the actual definition of $s_{ij}$.

## 4 Kleinberg-Tardos Matching Game

In this section, we consider the matching game studied by Kleinberg and Tardos [29] (KT-model). As a main result, Kleinberg and Tardos give a combinatorial characterization of stable outcomes which are balanced. Here, we present an economic characterization. Note that though the graph is a general network and not necessarily a bipartite graph, the capacity constraints on the agents are one. Thus the approach here, which is a special case of our more general approach in the rest of the paper, is much simpler and more intuitive for this setting.

Since the parameters and definitions are much simpler in matching game, we re-state them for the sake of better intuition. We model market structure as a network $G = (V, E)$ and each edge $e = (i,j) \in E(G)$ has value $w_e$ which can be divided unequally among $i$ and $j$ and each agent $i$ is allowed to participate in at most one contract. The outcome of the game is a matching $M$ and a vector $x \in R^N$, where $x_i$ is the income of agent $i$ (from its sole contract, if any). The outside option $\alpha_i$ is simply $\max_j \{w_{(i,j)} - x_j : (i,j) \notin M\}$; we take the maximum to be 0 over an empty set. An outcome is stable if for all $i$, $x_i \geq \alpha_i$, and is balanced if for all pair $(i,j) \in M$ we have $x_i - \alpha_i = x_j - \alpha_j$. For every $S \subseteq V(G)$, $\nu(S)$ is equal to the size of a maximum-weight matching $M_S$ in the induced subgraph $G[S]$ (as Kleinberg and Tardos [29] also point out it is very easy to see that a matching in a stable outcome must be a maximum-weight matching.). The definitions of core, the power of an agent with respect to another agent, prekernel, and excess are exactly those in Definitions 2.5- 2.6.

First we consider the core and simplify its first condition as follows: for all edges $e = (i,j)$, $x_i + x_j \geq w_e$. Note that though this condition is weaker, indeed it is equivalent: for a set $S$ for all edges $e = (i,j) \in M_S$, $x_i + x_j \geq w_e$ and thus $\sum_{i \in S} x_i \geq \nu(S)$. Next, we consider the power of agent $i$ with respect to agent $j$ and simplify it as follows: $s_{ij} = \max \{w_{ik} - x_i - x_k : (i,k) \in E(G), k \neq j\}$, we take the maximum to be $-x_i$ over an empty set. Again though this condition is weaker (say by taking all $S = \{i,k\} : (i,k) \in E(G))$, indeed it is equivalent since in any (non-empty) maximum matching $M_S$, each edge $e = (i',j') \in M_S$ contributes $w_e$ to $\nu(S)$ and at least $w_e$ to $\sum_{k \in S} x_k$ since $x_{i'} + x_{j'} \geq w_e$ and thus they cannot be of any help.

Now we are ready to prove two main theorems of this section relating stable and balanced with core and prekernel.

**Theorem 4.1.** *An outcome $(M, x)$ is stable if and only if payoff vector $x$ is in the core.*

**Proof.** We use the strong duality theorem and complementary slackness conditions to prove the lemma. First we prove if outcome $(M, x)$ is stable then $x$ is in the core. First we consider the second condition of the core. Since by definition $\sum_{i \in N} x_i = w_M$, where $w_M = \sum_{e \in M} w_e$, we only need to prove that $M$ is a maximum-weight matching. First we consider the following LP and its dual.

$$
\begin{array}{lll}
\text{minimize} & \sum_i x_i & \\
\text{subject to} & x_i + x_j \geq w_e & \forall (i,j) \in E(G) \\
& x_i \geq 0 & \forall i \in V(G)
\end{array} \quad \text{(LP3)}
$$

$$
\begin{array}{lll}
\text{maximize} & \sum_e w_e y_e & \\
\text{subject to} & \sum_{j : e=(i,j) \in E(G)} y_e \leq 1 & \forall i \in V(G) \\
& y_e \geq 0 & \forall e \in E(G)
\end{array} \quad \text{(LP4)}
$$

Note that $x$ and $M$ (i.e., $y_e = 1$ if $e \in M$, and $y_e = 0$ otherwise) are feasible solutions for LP3 and LP4. Since $\nu(V(G)) = \sum_{k \in V(G)} x_k = w_M$, by the strong duality theorem we can conclude that $M$ is a maximum-weight matching. Now we consider the aforementioned first simplified condition of the core.

Consider an edge $e = (i, j)$. For $e \in M$ we have $x_i + x_j = w_e$ by the definition of an outcome. For $e \notin M$, by the definition of an outside option, $\alpha_i \geq w_e - x_j$, and by the definition of stable, $x_i \geq \alpha_i$, which results in $x_i + x_j \geq 1$ as desired.

Next we prove there is a stable solution $(M, x)$ with respect to any vector $x$ in core. We prove this by showing a maximum matching $M$ for which for all $e = (i, j) \in M$, $x_i + x_j = w_e$. Note that $x$ is feasible solution for the LP3. Consider any maximum-weight matching $M$ and set $y_e = 1$ if $e \in M$, and $y_e = 0$ otherwise. Thus $y$ is a feasible solution for the dual LP4. Since $\nu(V(G)) = \sum_{k \in V(G)} x_k = |M|$, by the strong duality theorem $x$ and $y$ are optimum solutions of LP1 and its dual LP4. By complementary slackness conditions $y_e(x_i + x_j - w_e) = 0$, for each $e = (i, j)$. It means if $y_e > 0$, i.e., $e \in M$, then $x_i + x_j = w_e$, as desired. □

Indeed the proof of Theorem 4.1 gives a characterization of graphs with a non-empty set of stable outcomes.

**Corollary 4.2.** *A graph has non-empty core (and thus has a stable outcome) if and only if LP3 has integrality gap 1 for finding a maximum-weight matching.*

Note that it is well-known that LP3 has integrality gap 1 for bipartite graphs and thus for bipartite graphs always we have non-empty cores. On the other hand, it is well-known that if we add the following condition $\sum_{e \in G[S]} y_e \geq \lceil \frac{|S|-1}{2} \rceil$ for all subsets $S \subseteq V(G)$, then the integrality gap of LP3 is 1 for all graphs [38].

Finally, we prove the following theorem.

**Theorem 4.3.** *An outcome $(M, x)$ is stable and balanced if and only if the payoff vector $x$ is in the intersection of core and prekernel.*

**Proof.** By Theorem 4.1, we know that an outcome $(M, x)$ is stable if and only if vector $x$ is in the core. In addition, according to its proof, we can always construct a maximum matching $M$ corresponding to a vector $x$ in the core. We now prove if $(M, x)$ is balanced then $x$ is in the prekernel set. For two agents $i$ and $j$ we consider two cases.

1. $(i, j) \in M$: Indeed the same $k \neq j$ which maximizes $w_{ik} - x_i - x_k$ in the aforementioned simplified definition of $s_{ij}$ should maximize $\alpha_i$ too. Similarly, the same $k' \neq i$ which maximizes $w_{jk'} - x_j - x_{k'}$ in the aforementioned simplified definition of $s_{ji}$ should maximize $\alpha_j$ too. Thus the balanced condition $x_i - \alpha_i = x_j - \alpha_j$ implies $x_i - w_{ik} + x_k = x_j - w_{jk'} + x_{k'}$ which implies $s_{ij} = s_{ji}$.

2. $(i, j) \notin M$: Now if there is a $k$ such that $(i, k) \in M$, then $s_{ij} \geq w_{ik} - x_i - x_k = 0$. Since $x$ is in the core, $s_{ij} \leq 0$ and thus $s_{ij} = 0$. If there is no such $k$, $x_i = 0$. Consider any edge $(i, k) \in E(G)$, $k \neq j$. If there is no such $k$, then $s_{ij} = 0$ by the definition. Otherwise since $x_i = 0$, $x_k = w_{ik}$. Thus $s_{ij} \geq w_{ik} - x_i - x_k = 0$, which again implies $s_{ij} = 0$ since $x$ is in the core. Thus $s_{ij} = 0$. Similarly $s_{ji} = 0$ and thus $s_{ij} = s_{ji}$ as desired.

Finally, we prove if $x$ is in the prekernel then $x$ with its corresponding matching $M$ is balanced. By the definition of prekernel, for every edge $i, j \in V(G)$ we know that $s_{ij} = s_{ji}$. For an edge $(i, j) \in M$, by the simplified definition $s_{ij}$, $s_{ij} = \alpha_i - x_i$. Similarly $s_{ji} = \alpha_j - x_j$. Thus $s_{ij} = s_{ji}$ implies $x_i - \alpha_i = x_j - \alpha_j$ and thus $(M, x)$ in a balanced outcome. □

Kleinberg and Tardos [29] via a combinatorial proof establish the following main theorem of their paper.

**Theorem 4.4.** *If a graph $G$ has a stable outcome, then it has a balanced outcome, and the set of all balanced outcomes can be constructed in polynomial time.*

Indeed the proof of Theorem 4.4 follows simply from Theorem 4.3 and the known economic fact that if the core is non-empty (i.e., there is a stable outcome), the core intersection prekernel is non-empty (in this case the nucleolus is in the intersection of core and prekernel [41]). Constructibility simply follows because of constructibility of core intersection prekernel [21, 35] for this special case of a cooperative game.

To compute a point in the prekernel, Faigle et al. [21] only need an oracle that computes $s_{ij}(\mathbf{x})$ given any vector $\mathbf{x}$ (which is trivial in our case). The approach is a local search that converges to a point in the prekernel. They begin with a point $\mathbf{x}$ in the core, and so long as there are players $i, j$ for whom $s_{ij}(x) < s_{ji}(x)$, a transfer of $\frac{s_{ji}(x) - s_{ij}(x)}{2}$ is made from $i$ to $j$. This idea is generally attributed to Maschler, but Stearns [47] was the first one to propose a transfer scheme with provable convergence. His scheme always performs a transfer involving players $i, j$ with the minimal $s_{ij}$, and among all such possibilities, he picks a pair which maximizes the transfer amount. To prove convergence, one strategy is to show the total amount of transfers is bounded—this is done via a potential function approach. Some more work is required to show the local search procedure converges to a point in the prekernel. The number of steps, though, need not be polynomially bounded (or even finite). To guarantee this property, Faigle et al. [21] suggest alternating between transfer steps and a linear programming-based move. An LP move comes after each $O(n^2)$ transfers, which removes forever at least one pair from the set of players involved in the transfers.

Meinhardt [35] gives an LP-based algorithm that can find (almost) the entire prekernel in certain cases including the games we consider. They show how to find the prekernel via a sequence of LPs. In other words, their main contribution is a way to compress the transfer phases between the LP moves of the previous algorithm. He also presents a single linear program for finding one point of the prekernel.

## 4.1 Local dynamics for finding a balanced solution

This section proves Theorem 2.11. Stearns proposes a convergent scheme to find the prekernel in certain classes of cooperation games. Recall that $s_{ij} := min_{S \subseteq N \setminus \{j\}, i \in S}\{x(S) - \nu(S)\}$ for any two vertices $i$ and $j$.

**Theorem 4.5** ([47]). *Provided that the $s_{ij}$ vector is computable for a given bargaining game, there is a local scheme whose output converges to a point in the intersection of the prekernel and the least-core.*

Indeed, Faigle et al. [20] improve on this algorithm and make it run in polynomial time. We show how to apply this convergent scheme to our constrained bipartite bargaining game, in effect reproducing the result of [1]. The main task is showing that we can compute the $s_{ij}$ vector. Unfortunately, as in [1], we are unable to prove that the resulting dynamics converge in polytime.

**Lemma 4.6.** *Given a graph $G$, the degree upper bounds $c_v$, vertices $i, j$, and a vector $x$ in the core of the game, we can compute $s_{ij}$ in polynomial time for the matching game, as well as for the constrained bipartite bargaining game.*

Before starting to prove this lemma, we notice that Theorem 2.11 is immediate from Lemma 4.6 and Theorem 4.5.[8] The resulting algorithm is quite similar to what Azar et al. [1] recently suggested for this problem. One advantage of our result, though, is that we do not need to have a maximum-weight matching to carry out the procedure. Unlike the recent work, our algorithm is able to find the profit vector $x_i$ for all the agents without the knowledge of any matching. However, one should note that producing the actual contracts is an inevitable part of describing the outcome of the bargaining game.

---

[8]The assumption that $x$ is in the core does not hurt the generality of the approach since it is easy to find a core solution. There are even local dynamics finding it. In addition, the convergence algorithm has the property that it always gives a core solution to the $s_{ij}$ subroutine.

**Proof of Lemma 4.6.** We first prove the result for the matching game. The algorithm computes the value of $x(S) - \nu(S)$ for all sets $S : i \in S, j \notin S$ such that $|S| \leq 2$, and outputs their minimum value.

Let $\theta$ denote the output of our solution, and let $S^*$ be a *minimal* optimal set in the definition of $s_{ij}$. Clearly, $s_{ij} \leq \theta$. Let $E' \subseteq E$ denote a maximum matching of $S$. Since $x$ is in the core, we have $x_u + x_v - w_{uv} \geq 0$ for any edge $(u, v) \in E'$. Thus we can remove any edges not incident on $i$ from $E'$ without increasing the value of the set. Suppose we end up with a matching $E''$. There is at most one edge in $E''$. Let $S''$ be the endpoints of this edge if there is any, or the vertex $i$ otherwise. We know $\theta \leq x(S'') - \nu(S'')$ since $|S''| \leq 2$.

Now we provide the proof for the case of the constrained bipartite game: suppose the vertices are partitioned into $V_1$ and $V_2$ and degree bound of all the vertices in $V_2$ is one. A similar argument shows it is sufficient to consider all the connected graphs that have only one vertex from $V_1$. We look at an optimal $f$-factor in an optimal set $S$. Then we only consider the connected component containing $i$ since all the other components can be removed without increasing the value of the set; this follows from membership of $x$ in the core.

These structures do not have a constant size, so simple enumeration does not give a polynomial time bound.

Suppose $i \in V_1$. Sort all the other vertices $j'$ in increasing order of $x(\{j'\}) - w_{ij'}$. We can ignore any vertex $j'$ for which this parameter is not negative. Among the remaining vertices, we pick the best $c_i$ vertices or simply all the vertices if $c_i$ is larger than their count. It is easy to see this gives the best solution.

If $i \in V_2$, we guess a vertex $j' \in V_1$ as the "center" of the star. The rest of the argument is similar to the above paragraph except that we have to include vertex $i$ in the star. Another alternative is to consider the singleton set $\{i\}$. This shows that we can compute $s_{ij}$ in polynomial time for the constrained bipartite game. □

## 5 Finding the Nucleolus

In this section we propose an algorithm to find the nucleolus for our constrained bipartite bargaining game in polynomial time. Previous works [28, 46, 20] show how to compute the nucleolus via an iterated LP-based algorithm, and we also adopt this approach. The main complication in applying this algorithm is that the natural LP does not have polynomial size, and hence we must prune it without sacrificing the correctness of the algorithm.

We describe this general approach as it applies to our setting. As the core is not empty in our constrained bipartite bargaining game (see Section 3), the nucleolus will be in the core. Hence, our task is to search for a core outcome $\{x_i\}$ which maximizes the lexicographically-sorted sequence of set excesses $\epsilon = (\epsilon_{S_1}, \ldots, \epsilon_{S_{2N}})$; see Definition 2.7. We proceed iteratively. In each iteration, we search for the "next" element $\epsilon_{S_i}$ of $\epsilon$ by solving a linear program whose objective defines $\epsilon_{S_i}$. In doing so, we must be careful to constrain certain variables appropriately such that the computations of previous iterations carry through. As there are exponentially many elements of the vector $\epsilon$, we cannot simply introduce an equality for each element that we wish to fix. Rather, we show how to construct a polynomially-sized set of "representatives" and argue that these suffice to describe the excesses of sets fixed in all previous iterations.

## 5.1 LP formulation

We begin by introducing a linear program that defines the excesses. This LP will have exponential size; we will show how to prune it in later sections. Consider the first iteration. In this iteration, we want to find a core outcome with the largest smallest excess. That is, we must maximize $\alpha_k$, the excess, subject to (1) $\sum_{i \in T} x_i \geq \nu(T) + \alpha_k$ for all non-empty $T \subset N$ (all sets have excess at least $\alpha_k$), and (2) $\sum_{i \in N} x_i = \nu(N)$ (**x** is in the core).

After solving this LP, we will have fixed some of the excesses in our solution. In all following iterations, we must make sure that the excesses fixed in this iteration remain unchanged. In order to do this, we introduce additional inequalities. Let $\mathcal{F}_k$ denote the set of sets for which $\epsilon_{S_i} = \alpha_k$, and $\mathcal{F}_k^* = \bigcup_{i=1}^{k} \mathcal{F}_i$. Clearly, the excess of sets in $\mathcal{F}_k^*$ must remain constant. This requirement implies conditions on the values of **x** which in turn may force the excesses of additional sets to be fixed. For example, given two sets $T_1$ and $T_2$ in $\mathcal{F}_k^*$ and a third set $T_3$ for which $\epsilon_{T_3} = \epsilon_{T_1} + \epsilon_{T_2}$, then $T_3$ has excess fixed to $2\alpha_k$ although $T_3$ is not in $\mathcal{F}_k^*$. We let $\mathcal{F}_k^+$ denote the set of sets whose excess is fixed by iteration $k$. Note that $\mathcal{F}_k^+$ is a superset of $\mathcal{F}_k^*$ as $\mathcal{F}_k^+$ contains any set which has been fixed using fixed values for sets in $\mathcal{F}_k^*$. In order to find the next excess values, we must now solve the following LP.

$$
\begin{array}{ll}
\text{maximize} & \alpha_k \\
\text{subject to} & \sum_{i \in T} x_i = \nu(T) + \alpha_j \qquad \forall j < k, \forall T \in \mathcal{F}_j \\
& \sum_{i \in T} x_i \geq \nu(T) + \alpha_k \quad \forall T \subseteq N, T \notin \mathcal{F}_{k-1}^+, T \neq N, T \neq \emptyset \\
& \sum_{i \in N} x_i = \nu(N) \\
& x_i \geq 0 \qquad \forall i \in N,
\end{array}
\quad \text{(LP5)}
$$

where the first inequality guarantees that the sets in $\mathcal{F}_k^+$ remain fixed, and the second inequality is only written for unfixed sets (note that fixed sets might really violate this inequality).

## 5.2 Solving the LP

Our proposed algorithm is to iteratively solve a sequence of LPs defined by LP5. As has been observed previously, this algorithm will have at most $n$ iterations [20]. In each iteration, at least one inequality goes tight, and as the core consists of $n$ variables, $n$ equalities suffice to define it.

The main technical difficulty we face is that the size of our LP5 is exponential. In this section, we show how to prune our LP in each iteration to get an LP of polynomial size. Consider an arbitrary set of optimal contracts $M$ in our network. Note that as we discussed in Section 3, any outcome in the core (including nucleolus) can construct a stable solution with this optimal $M$. Note that, as we are considering the constrained bipartite bargaining game, the set of edges in $M$ is simply a union of stars, each of which has its center vertex in $V_1$ and the remaining vertices in $V_2$. We denote these stars by $R_1, \ldots, R_m$, and their centers by $r_1, \ldots, r_m$. We will call edges in $M$ "green" edges and the remaining edges in the graph "black" edges. Recall that the vertices in $V_2$ all have capacity one. We now prune LP5 by eliminating all inequalities involving any set $T$ for which

1. $T$ has more than 1 vertex in $V_1$: $|T \cap V_1| > 1$ (Lemma 5.1);

2. some vertex $i \in T \cap V_1$ has degree greater than $c_i + 1$ (Lemma 5.2);

3. some vertex $i \in T \cap V_1$ has degree equal to $c_i + 1$ and $\geq 3$ black edges, or degree at most $c_i$ and $\geq 2$ black edges (Lemma 5.3); or

4. some vertex $i \in T \cap V_1$ has at most $c_i - 2$ green edges (Lemma 5.4).

Before giving a precise proof of the above claim, let us develop some intuition into the idea. Suppose we want to prove that we could ignore a set $T$. We look for some sets $T_1, ..., T_m$ and prove $\epsilon(T_i) \leq \epsilon(T)$, for every $1 \leq i \leq m$. So every $T_i$ has been fixed before $T$ (or at the same time). Now if we prove we can write $\epsilon(T)$ as $\lambda_1 \epsilon(T_1) + ... + \lambda_m \epsilon(T_m) + \lambda$, then we can show it causes no harm to ignore $T$ in LP5. In fact, if all the sets $T_j : 1 \leq j \leq m$ have been fixed at some time $t \leq k$, then we have $T \in \mathcal{F}_t^+$.

Because of properties one and two above, we only need to consider sets which have one vertex in $V_1$ and at most $c_i + 1$ vertex in $V_2$. By property two, we know that among these sets we should consider those which have at most two black edges (that we can try all). Note that however still the number of such sets can be exponential, since they can have any subset of green edges plus zero, one, or two black edges. Property four indeed restricts the number of excluded green edges to at most two (that again we can try all). As a result, only a polynomial number of sets remain. Therefore, the linear program can be solved in polynomial time and we are done with the proof of Theorem 2.12.

**Lemma 5.1.** *In solving LP5 we can ignore any set $T$ which has more than one vertex in $V_1$.*

**Proof.** Consider a set $T \in \mathcal{F}_k$, and suppose it has more than one vertex in $V_1$, say $r_1, ..., r_p$. Run LP1 for the graph induced by $T$ and find a solution corresponding to $\nu(T)$. Let $x_i^T$ denote the sum of the weights of green edges connected to $r_i$ in the optimum solution of LP1 for $T$. Also call $r_i$ with all its green edges $R_i^T$. We know that $\nu(T) = x_1^T + \cdots + x_p^T$ and $\nu(R_i^T) = x_i^T$. Therefore, we have $\epsilon(T) = \epsilon(R_1^T) + \cdots + \epsilon(R_p^T)$. Nonnegativity of the excess values implies that $R_i^T$ has been fixed before $T$ or at the same time for every $i : 1 \leq i \leq p$. So if we ignore $T$ we could solve LP5 and we have $T \in \mathcal{F}_k^+$. In fact if all $R_i^T$ sets have been fixed at time $t \leq k$, we can compute $\epsilon(T)$ based on equation $\epsilon(T) = \epsilon(R_1^T) + \cdots + \epsilon(R_p^T)$. Therefore, $T \in \mathcal{F}_t^+$. □

**Lemma 5.2.** *In LP5 we can ignore any set $T$ with exactly one vertex $i$ in $V_1$ whose degree in $T$ is larger than $c_i + 1$.*

**Proof.** Consider a set $T \in \mathcal{F}_k$ with one $V_1$ vertex, say $i$, that has edges to $m > c_i + 1$ vertices of $T \cap V_2$, say $j_1, \ldots, j_m$, as shown in Figure 3. Assume $w_{i,j_k} \geq w_{i,j_{k'}}$ for every $1 \leq k < k' \leq m$. We know that $\nu(T) = \nu(T \setminus \{j_l\}) = \sum_{k=1}^{c_i} w_{i,j_k}$ for every $l > c_i$. Therefore, for every $l > c_i$, $\epsilon(T) \geq \epsilon(T \setminus \{j_l\})$. Let $T' = T \setminus \{j_{c_i+1}, \ldots, j_m\}$. It is clear that $\nu(T') = \nu(T) = \sum_{k=1}^{c_i} w_{i,j_k}$ and so $\epsilon(T') \leq \epsilon(T)$. Thus, the values of $T \setminus \{j_l\}$ and $T'$ have been fixed before $T$ or at the same time. Suppose sets $T \setminus \{j_l\}$ for all $l : l > c_i$ and $T'$ have been fixed at time $t \leq k$. It is now sufficient to show that $\epsilon(T)$ can be inferred from looking at $\epsilon(T \setminus \{j_l\})$ and $\epsilon(T')$ values to prove $T \in F_t^+$. By adding up all the equations $\epsilon(T \setminus \{j_l\}) = \sum_{i \in T \setminus \{j_l\}} x_i - \nu(T)$ for $l > c_i$, we get $\sum_{l=c_i+1}^{m} \epsilon(T \setminus \{j_l\}) = (m - c_i) \sum_{i \in T} x_i - \sum_{l=c_i+1}^{m} x_{j_l} - (m - c_i)\nu(T)$. In fact, we can write $\epsilon(T)$ as $\epsilon(T) = \sum_{i \in T} x_i - \nu(T) = \frac{1}{m - c_i - 1} \left( \sum_{l=c_i+1}^{m} \epsilon(T - \{j_l\}) - \epsilon(T') \right)$. Therefore we can find $\epsilon(T)$ based on already-fixed values of $\epsilon(T - \{j_l\})$ for every $l > c_i$ and $\epsilon(T')$. □

**Lemma 5.3.** *In solving LP5 we can ignore any set $T$ with exactly one vertex $i$ in $V_1$ if*

- *degree of $i$ in $T$ is at least $c_i + 1$ and it has at least 3 black edges, or*
- *degree of $i$ in $T$ is at most $c_i$ and it has at least 2 black edges.*

**Proof.** Consider a set $T \in \mathcal{F}_k$ with $\{i\} = T \cap V_1$ fitting the description in the statement of the lemma. We can prove the statement using Lemma 5.2 if degree of $i$ in $T$ is greater than $c_i + 1$. Suppose thus that the degree of $i$ in $T$ is $d_i^T \leq c_i + 1$, and $i$ has $m_1$ black edges to vertices $j_1, \ldots, j_{m_1}$ in $T$, and $i$ has $m_2$

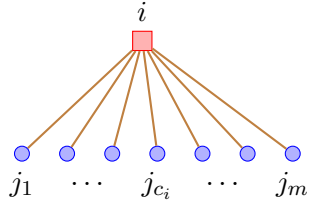

Figure 3: Degree of $i$ is greater that $c_i + 1$

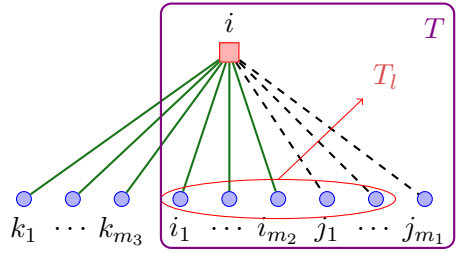

Figure 4: Vertex $i$ with at least 2 black edges

green edges to vertices $i_1, \ldots, i_{m_2}$ in $T$, and $i$ has $m_3$ green edges to vertices $k_1, \ldots, k_{m_3}$ in $V_2 \setminus T$; refer to Figure 4. Let $\nu(T)$ be equal to $\sum_{t \in T_l} w_{it}$. Note that if $d_i^T \leq c_i$ then $T_l = T \setminus \{i\}$. There are at least two black edges in $T_l$, because $m_1 \geq 2$ (see the statement of the lemma). On the other hand, if $d_i^T = c_i + 1$ then $T_l = T \setminus \{i, l\}$ where $l \in T \cap V_2$ has the minimum weight edge to $i$ among all vertices of $T \cap V_2$. We know that $m_1 \geq 3$ in this case. So there are at least two black edges in $T_l$ in this case too. Without loss of generality, call these two black edges $(i, j_1)$ and $(i, j_2)$. Now we have two cases to consider:

**Vertex $i$ is used up to its capacity:** In this case, we know that $m_2 + m_3 = c_i$ and $m_1 + m_2 = d_i^T$, which imply $m_3 = m_1 + c_i - d_i^T$. Therefore $m_3 \geq 2$ for $d_i^T \leq c_i$. But if $d_i^T = c_i + 1$, we have $m_1 \geq 3$ which gives $m_3 \geq 2$. So, we have $m_3 \geq 2$ in both cases. Let $T' = T \cup \{k_1\} \setminus \{j_1\}$. We prove $\epsilon(T') \leq \epsilon(T)$. First, note that $\nu(T) = \sum_{h \in T_l} w_{ih}$. On the other hand, we have a solution $\sum_{h \in T_l - \{j_1\} \cup \{k_1\}} w_{ih}$ for $\nu(T')$ which implies $\nu(T') \geq \sum_{h \in T_l \cup \{k_1\} \setminus \{j_1\}} w_{ih}$. So we have

$$\epsilon(T) - \epsilon(T') = x_{j_1} - x_{k_1} - \nu(T) + \nu(T')$$
$$\geq x_{j_1} - x_{k_1} - w_{ij_1} + w_{ik_1}.$$

We know that $(i, k_1)$ is a green edge and $(i, j_1)$ is a black edge, thus, $i$ prefers to have an agreement with $k_1$ instead of $j_1$. Hence, we conclude $w_{ik_1} - x_{k_1} \geq w_{ij_1} - x_{j_1}$ implying $\epsilon(T') \leq \epsilon(T)$. The same arguments hold for sets $T'' = T \cup \{k_2\} \setminus \{j_2\}$ and $T^* = T \cup \{k_1, k_2\} \setminus \{j_1, j_2\}$, and we can prove $\epsilon(T'') \leq \epsilon(T)$ and $\epsilon(T^*) \leq \epsilon(T)$. Thus, all sets $T'$, $T''$ and $T^*$ have been fixed before $T$ at some time $t \leq k$. We now show we can compute $\epsilon(T)$ based on $\epsilon(T')$, $\epsilon(T'')$ and $\epsilon(T^*)$. Note that

$$\epsilon(T') + \epsilon(T'') - \epsilon(T^*) + \nu(T') + \nu(T'') - \nu(T^*) - \nu(T) =$$
$$\sum_{h \in T'} x_h + \sum_{h \in T''} x_h - \sum_{h \in T^*} x_h - \nu(T) = \sum_{l \in T} x_l - \nu(T) = \epsilon(T).$$

Therefore, $T \in F_t^+$ can be ignored in the LP computation.

**Vertex $i$ does not exhaust its capacity:** Consider $i$ with all its green edges in the graph. Let us name this set $R_i$. Because $i$ is not saturated, we have $\nu(R_i \cup \{j_1\}) = \nu(R_i) + w_{ij_1}$. On the other hand, from the core constraints, we have $\sum_{l \in R_i \cup \{j_1\}} x_l \geq \nu(R_i) + w_{ij_1} = \sum_{l \in R_i} x_l + w_{ij_1}$,[9] and so we know $x_{j_1} \geq w_{ij_1}$. Let $T' = T \setminus \{j_1\}$. It is clear that $\nu(T) = \sum_{h \in T_l} w_{ih}$. Since $j_1 \in T_l$ we can conclude that $\nu(T') \geq \sum_{h \in T_l} w_{ih} - w_{ij_1} = \nu(T) - w_{ij_1}$. Therefore $\epsilon(T') - \epsilon(T) \leq \sum_{h \in T'} x_h - \sum_{h \in T} x_h + w_{ij_1} = w_{ij1} - x_{j_1} \leq 0$, which means $\epsilon(T') \leq \epsilon(T)$. Assume $T'' = T - \{j_2\}$ and $T^* = \{j_1, j_2\}$. With the same argument, we can prove $\epsilon(T'') \leq \epsilon(T)$ and $\epsilon(T^*) \leq \epsilon(T)$. So all sets $T'$, $T''$ and $T^*$ has been fixed before $T$ at some time $t \leq k$. To compute $\epsilon(T)$ based on $\epsilon(T')$, $\epsilon(T'')$ and $\epsilon(T^*)$, notice that

$$\epsilon(T') + \epsilon(T'') - \epsilon(T^*) + \nu(T') + \nu(T'') - \nu(T^*) - \nu(T) =$$
$$\sum_{h \in T'} x_h + \sum_{h \in T''} x_h - \sum_{h \in T^*} x_h - \nu(T) = \sum_{l \in T} x_l - \nu(T) = \epsilon(T).$$

This concludes the proof of this case, as well as the lemma. □

**Lemma 5.4.** *To solve LP5, it does not cause any harm to ignore a set $T$ with exactly one vertex in $V_1$ excluding at least 2 green edges.*

**Proof.** Consider a set $T \in \mathcal{F}_k$ with $\{i\} = T \cap V_1$ fitting the description of the lemma. We can apply Lemma 5.2 if the degree of vertex $i$ is greater that $c_i + 1$. Moreover, if the degree of vertex $i$ is $c_i + 1$ or $c_i$ and excludes at least 2 green edges, Lemma 5.3 can be invoked. So suppose the degree of vertex $i$ is at most $c_i - 1$. We first show that $x_j \leq w_{ij}$ for every green edge $(i, j)$. Consider vertex $i$ with all its green edges and call this set $R$. We know from the discussion of Section 3.2 that $\sum_{l \in R} x_l = \nu(R)$. If we remove a green edge $(i, j)$ from $R$, we have $\sum_{l \in R \setminus \{j\}} x_l \geq \nu(R \setminus \{j\}) = \nu(R) - w_{ij}$. So $x_j \leq w_{ij}$. Consider the set $T$. Because $d_i^T \leq c_i - 1$, $\nu(T) = \sum_{l \in T \cap V_2} w_{il}$. Also there are at least two green edges $(i, j_1)$ and $(i, j_2)$ which are not in $T$ as shown in Figure 5. Now there are two possibilities:

- Degree of vertex $i$ it less that $c_i - 1$: Consider $T_1 = T \cup \{j_1\}$ and $T_2 = T \cup \{j_2\}$. Degree of $i$ in $T_1$ and $T_2$ is at most $c_i - 1$, so we have $\nu(T_1) = \sum_{l \in T_1 \cap V_2} w_{il}$ and $\nu(T_1) = \sum_{l \in T_2 \cap V_2} w_{il}$. Our goal is to prove $\epsilon(T_q) \leq \epsilon(T)$ for $q = 1, 2$.

$$\epsilon(T_q) - \epsilon(T) = \sum_{l \in T_q} x_l - \sum_{l \in T_q \cap V_2} w_{il} - \sum_{l \in T} x_l + \sum_{l \in T \cap V_2} w_{il} = x_{j_q} - w_{ij_q}.$$

But we know that $(i, j_q)$ is a green edge and we have proved that $x_{j_q} \leq w_{ij_q}$ in this case, which means that $\epsilon(T_q) \leq \epsilon(T)$. Degree of $i$ is at most $c_i$ in the set $T_3 = T \cup \{j_1, j_2\}$. Therefore, the same arguments hold for the set $T_3$ and we can prove $\epsilon(T_3) \leq \epsilon(T)$. Therefore, $T_q$ has been fixed before $T$ for $1 \leq q \leq 3$. Suppose all $T_q$ has been fixed at time $t \leq k$. We also have

$$\epsilon(T) = \epsilon(T_1) + \epsilon(T_2) - \epsilon(T_3) + \nu(T_1) + \nu(T_2) - \nu(T_3) + \nu(T).$$

Hence, the value of $\epsilon(T)$ has been fixed at a time $t$ when all values $\epsilon(T_1)$, $\epsilon(T_2)$ and $\epsilon(T_3)$ have been fixed. Thus, $T \in \mathcal{F}_t^+$.

- Degree of vertex $i$ is exactly $c_i - 1$. Because $i$ excludes at least 2 green edges $(i, j_1)$ and $(i, j_2)$, it has at least one black edge to $b$. Note that if $i$ has more than one black edges, we can use Lemma 5.2 to solve the problem. So assume $i$ has exactly one black edge; see Figure 5. Observe that if it has no black edges, it should have all its green edges in $T$ except at most one, which is in contradiction with the lemma's statement. Consider the sets $T_1 = T \cup \{j_1\} \setminus \{b\}$, $T_2 = T \cup \{j_2\}$, and $T_3 = T \cup \{j_1, j_2\} \setminus \{b\}$. Because $(i, j_2)$ is a green edge, we have $x_{j_2} \leq w_{ij_2}$, which gives $\epsilon(T_2) \leq \epsilon(T)$. On the other hand, we know that $(i, b)$ is a black edge and $(i, j_1)$ is a green edge, implying that $i$ prefers to have an agreement with $j_1$ instead of $b$. Thus, $w_{ij_1} - x_{j_1} \geq w_{ib} - x_b$. Let us now compute values $\epsilon(T_1) - \epsilon(T)$ and $\epsilon(T_3) - \epsilon(T_2)$:

$$\epsilon(T_1) - \epsilon(T) = x_{j_1} - w_{ij_1} - x_b + w_{ib} \leq 0$$
$$\epsilon(T_3) - \epsilon(T_2) = x_{j_1} - w_{ij_1} - x_b + w_{ib} \leq 0.$$

---

[9]For the equality, see Section 3.

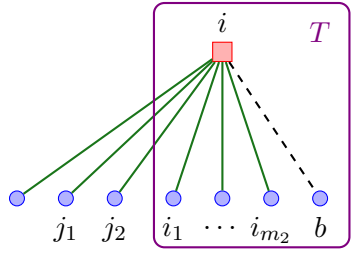

Figure 5: Vertex $i$ without at least 2 green edges

We conclude that all these sets have been fixed before $T$. Now we show that if all these sets have been fixed at some time $t \leq k$, then $T \in \mathcal{F}_t^+$. To prove this result, we write $\epsilon(T)$ in terms of values $\epsilon(T_1), \epsilon(T_2)$, and $\epsilon(T_3)$. First let us write all excesses in this form:

$$\epsilon(T) = \sum_{w \in T} x_w - \nu(T)$$
$$\epsilon(T_1) = \sum_{w \in T} x_w + x_{j_1} - x_b - \nu(T_1)$$
$$\epsilon(T_2) = \sum_{w \in T} x_w + x_{j_2} - \nu(T_2)$$
$$\epsilon(T_3) = \sum_{w \in T} x_w + x_{j_1} + x_{j_2} - x_b - \nu(T_3),$$

to derive $\epsilon(T) = \epsilon(T_1) + \epsilon(T_2) - \epsilon(T_3) + \nu(T_1) + \nu(T_2) - \nu(T_3) - \nu(T)$. This finishes the proof of $T \in \mathcal{F}_t^+$. □

**Acknowledgements.** The authors would like to acknowledge Éva Tardos for suggesting possible connections with the cooperative game theory literature.